# Coulomb gap induced by electronic correlation and enlarged superconducting gap in laterally confined Pb islands grown on SrTiO$_3$

Yonghao Yuan[1,2], Xintong Wang[1,2], Canli Song[1,2], Lili Wang[1,2], Ke He[1,2], Xucun Ma[1,2], Hong Yao[1,2,3], Wei Li[1,2]* and Qi-Kun Xue[1,2,4]*

[1]*State Key Laboratory of Low-Dimensional Quantum Physics, Department of Physics, Tsinghua University, Beijing 100084, China*
[2]*Collaborative Innovation Center of Quantum Matter, Beijing 100084, China*
[3]*Institute for Advanced Study, Tsinghua University, Beijing 100084, China*
[4]*Beijing Academy of Quantum Information Sciences, Beijing 100193, China*
*To whom correspondence should be addressed: weili83@tsinghua.edu.cn; qkxue@mail.tsinghua.edu.cn

We report high-resolution scanning tunneling microscopy (STM) study of nano-sized Pb islands grown on SrTiO$_3$, where three distinct types of gaps with different energy scales are revealed. At low temperature, an enlarged superconducting gap ($\Delta_s$) emerges while there is no enhancement in superconducting transition temperature ($T_c$), giving rise to a larger BCS ratio $2\Delta_s/k_BT_c \sim 6.22$. The strong coupling here may originate from the electron-phonon coupling on the metal-oxide interface. As the superconducting gap is suppressed under applied magnetic field or at elevated temperature, Coulomb gap and pseudogap appear, respectively. The Coulomb gap is sensitive to the lateral size of Pb islands, indicating that quantum size effect is able to influence electronic correlation, which is usually ignored in low-dimensional superconductivity. Our experimental results shall shed important light on the interplay between quantum size effect and correlations in nano-sized superconductors.

Electronic properties in quantum materials may be qualitatively modified by reducing their size and/or dimensions, which have attracted extensive interests in condensed matter research. One intriguing discovery in size confined systems is the enhancement of superconductivity in nano-sized weak coupling superconducting films, such as Al, In and Sn [1,2]. The mechanism of enhanced superconductivity has been explained as surface phonon softening [3], which leads to the enhanced electron-phonon coupling based on Eliashberg theory [4,5]. In contrast, $T_c$ of strong coupling superconductors, like Pb, is hardly boosted by reducing their size. Actually, in two-dimensional Pb films grown on Si(111), $T_c$ shows oscillating decrease tendency with the decreasing of film thickness [6,7]. To obtain enhanced the surface phonon coupling, Pb nano-particles with higher surface area ratio are prepared, in which the $T_c$ presents a constant value (still lower than that of bulk Pb) with the size reduction until approaching the Anderson limit [8,9].

When their sizes are smaller than the Anderson limit (~ 10 nm), both the weak and strong coupling superconductors lose their superconductivity [10]. In this zero-dimensional limit, extreme electronic correlation dominates and induces Coulomb blockade phenomenon. Indeed, studies on nano-sized silver particles [11] and graphene nano-ribbons [12] have shown that Coulomb interaction is significantly enhanced during size reduction, which may even induce metal-insulator transition. Besides the electron-electron interaction, electron-phonon interaction is modified by size effect as well, manifesting as pseudogap and pseudopeak features observed in Pb islands [13]. Therefore, electronic properties of low dimensional superconductors are more complicated than expected, and we need to revisit the influences of both Coulomb and electron-phonon interactions in those systems.

It has been experimentally established that superconductivity can be significantly enhanced in monolayer FeSe film grown on SrTiO$_3$ [14], demonstrating that metal-oxide interface could be another key factor to boost $T_c$ [14-16]. Hence, by constructing a new system that combined all those factors together, it would provide an ideal platform to study their interplay with two-dimensional superconductivity.

Here we report on molecular beam epitaxy (MBE) growth and systematic STM studies of laterally confined Pb islands on SrTiO$_3$(001) (Pb/STO). Three distinct types of energy gaps have been observed, which are superconducting gap, Coulomb gap, and pseudogap. They manifest themselves in different energy scales. At low temperature, superconducting gap in Pb/STO is enlarged, with the gap size ($\Delta_s$) of ~ 2.0 meV, larger than 1.36 meV for bulk Pb. However, the $T_c$ in Pb/STO is not enhanced from the bulk value, giving rise to a larger BCS ratio $2\Delta_s/k_BT_c$ ~ 6.22 in the nano-sized Pb/STO. When the superconductivity is suppressed under applied magnetic field, Coulomb gap ($\Delta_C$ ~ 1 meV) induced by electronic interaction is clearly revealed. The value of Coulomb gap increases with reduction of the lateral size of Pb islands. As the lateral size approaches the Anderson limit, Coulomb blockade feature starts to appear. Finally, a pseudogap $\Delta_p$ with the magnitude of ~ 10 meV, at rather larger energy scale, is visible (even at 30 K), which may originate from the weaken of electron-phonon interaction [13].

Our experiments were conducted in a commercial ultra-high vacuum STM (Unisoku), which is connected to a MBE chamber for sample preparation. In MBE chamber, Nb-doped STO substrates were pre-treated at 1200 ℃ to obtain TiO$_2$-terminated surfaces. High-purity Pb (99.999%) was thermally evaporated from a standard Knudsen cell at 450 ℃. During the sample growth, the substrate was kept at room temperature. Our STM measurement was carried out at a base temperature of 400 mK. Vertical magnetic field up to 15 T can be applied. d$I$/d$V$ spectra were measured using standard lock-in method with 973.0 Hz bias modulations.

Figure 1a shows a typical STM topographic image of Pb islands grown on SrTiO$_3$ substrate. The growth follows Stranski-Krastanov mode (i.e., islands start to form after the growth of a wetting layer of Pb). Three types of Pb islands classified as type A, B and C are obtained. The type A islands are thicker than 15 nm, and exhibit similar superconducting gap to that of bulk Pb (see in Fig. 1b). The U-shaped superconducting gap is fitted well by Dynes model [17], and the fitted gap value $\Delta_A$ is 1.36 meV with scattering rate $\Gamma_A$ ~ 0.08 meV, consisting with that of its bulk material.

The heights of the type B Pb islands are 3 ~ 8 nm and their areas ranges from 1000 nm$^2$ to 100000 nm$^2$. The d$I$/d$V$ spectrum shows a larger superconducting gap but with lower coherence peaks (Fig. 1c). The fitting result suggests that the magnitude of the enlarged gap $\Delta_B$ is 2.58 meV and the corresponding scattering rate $\Gamma_B$ is 0.24 meV. The superconducting gap enlargement of Pb has not been reported yet, and we attribute this to the special substrate (STO) used here. The STO-Pb interface might provide additional electron-phonon coupling channel to enhance the superconducting pairing strength. We note that the fitting curve deviates from the data near the Fermi level ($E_F$), implying the existence of non-BCS behaviors in the type B islands. Type C islands are of the smallest size, usually smaller than 500 nm$^2$. The d$I$/d$V$ spectrum (Fig. 1d) shows several discrete peaks, which are signatures for Coulomb blockade [18]. Meanwhile, superconductivity here is absent due to approaching of the Anderson limit [10]. Since type B islands present the most intriguing electronic properties, we mainly focus on this type of Pb islands in our following discussion.

Superconducting gaps of type B Pb islands are strongly suppressed by externally applied magnetic field. Figure 2a exhibits d$I$/d$V$ spectra taken on an island with applying various perpendicular magnetic fields. The effective radius (*R*) of the island is ~ 32.8 nm, where *R* is estimated by the formula $R = (Area/\pi)^{1/2}$. The superconducting gap keeps shrinking with increasing of applied magnetic fields and persists even up to 0.75 T, indicating an enhancement of upper critical field of the Pb islands compared with its bulk value ~ 0.08 T. Remarkably, the gap feature is no longer sensitive to magnetic field from 1.5 T and such a residual gap is unchanged up to 15 T. The magnitude of this field insensitive gap ($\Delta_C$) is ~ 0.65 meV (see Fig. 2a and supplementary materials). We would like to emphasize that, even when we consider only the magnetic sensitive part as the superconducting gap in the spectrum, the size of superconducting gap ($\Delta_s$) would be $\Delta_B - \Delta_C$ = 1.93 meV, which is still larger than that of bulk Pb.

The residual gaps are V-shaped (see Fig. 2a and 2b) and the density of states (DOS) exhibit a linear dependence on energy near $E_F$. Theoretical work has proposed that in two-dimensional correlated systems, Coulomb gap may develop and the DOS of corresponding electrons near $E_F$ linearly depend on energy: $G \sim |E|$ [19]. Therefore, the residual gap here might be Coulomb gap induced by electronic correlation. Previous studies on Coulomb gap mainly focused on amorphous metals [20] and doped semimetals [21], in which variable-range-hopping of localized electrons with poor screening dominates the conductivity. By reducing the size of a Pb island to cross from two-dimensional to zero-dimensional limit, Coulomb gap becomes pronounced for the following reasons: (*i*) Coulomb interaction affects in longer range according to Lindhard screening theory [22]. (*ii*) The itinerant *s*-electrons have spill-out effect, which further weaken the screening effect on smaller islands [23].

To investigate the relationship between the Coulomb gap and the size of Pb island, a series of d*I*/d*V* spectra (Fig. 2b) are taken on various Pb islands. High magnetic fields are applied to suppress the superconducting gap and highlight the Coulomb gap. As expected, $\Delta_C$ becomes larger and deeper as *R* decreases, showing that Coulomb interaction is indeed affected by the size effect and is more pronounced in smaller islands. Based on the fitting of our results, the Coulomb gap size ($\Delta_C$) has a power-law dependence on *R*: $\Delta_C \sim 1/R^{0.353}$. $\Delta_C$ and *R* are plotted out in logarithmic scales in Fig. 2c, where the exponent 0.353 is obtained from the slope of the fitting curve. Similarly, as shown in Fig. 2d, the logarithmic zero energy density of states ($G_0$) obeys $\ln G_0 \sim -1/R^{1.91}$, where the exponent 1.91 is very close to 2, indicating $G_0$ decays exponentially with 1/*Area*. It is noteworthy that the spectrum on a small island with *R* = 23.1 nm shows a harder gap and several peak feature (denoted by cyan arrows in Fig. 2b) outside the gap, indicating that the system approaches Coulomb blockade regime (like type C islands) as the lateral size of island keeps decreasing. Pb islands were grown on $SrTiO_3$ substrates with different Nb doping levels (0.7%, 0.25% and 0.15%), while the data taken on those substrates obey the same fitting curve (see the data points with different color in Fig. 2b-d), demonstrating that the Coulomb gap is not sensitive to the details of substrates as well as their interfaces.

We further studied magnetic field response on a larger type B island (*R* ~ 83.7 nm). Since the Bean-Livingston barrier [24] is overcome, a magnetic vortex is observed. Fig. 3a shows zero-bias d*I*/d*V* maps taken under different magnetic fields. The vortex emerges at 0.18 T and becomes blurred at 0.27 T. A linear fit on the extracted zero bias conductance values away from the vortex suggests an upper critical field ($\mu_0H_{c2}$) of 0.30 T (see supplementary materials). $H_{c2}$ is determined by G-L coherence length $\xi_{GL}$: $\mu_0H_{c2} = \Phi_0/2\pi\xi_{GL}^2$, where $\Phi_0$ is magnetic flux quanta. Accordingly, the calculated coherence length $\xi_{GL}$ is 33 nm, smaller than that of bulk Pb value (83 nm). A series of d*I*/d*V* spectra are taken across a vortex at 0.20 T and shown in Fig. 3b. A zero-bias peak originating from the Caroli-de Gennes-Matricon bound states is clearly revealed in the vortex core, indicating that this two-dimensional superconducting system is in clean limit. In contrast, the superconducting Pb islands grown on Si (111) are always in dirty limit, and the bound states are absent in the vortex core [25].

Temperature dependent measurements show that the superconducting gap is almost filled at 7.0 K (Fig. 3c). The fitting result suggests that the $T_c$ is not higher than 7.3 K (see supplementary materials), which does not show significant enhancement compared with the $T_c$ of bulk Pb. Intriguingly, the unchanged $T_c$ associated with the enlarged superconducting gap $\Delta_s$ leads to a rather large BCS ratio ($2\Delta_s/k_BT_c$) ~ 6.22, close to the ratio in some cuprates [26]. The enlarged BCS ratio indicates that Pb, an inherent strong coupling superconductor, becomes even more strong-coupled when grown on STO. Theoretical work has pointed out that the BCS ratio may be increased through the mechanism of surface phonon softening [27]. In our case, electron-phonon coupling on metal-oxide interface provides another possible channel to enhance the coupling strength.

Normal state of Pb islands is further investigated by larger energy range d$I$/d$V$ spectra. As shown in Fig. 4a, a larger gap $\Delta_p$ with magnitude of ~ 10 meV is observed above $T_c$ and even exists at 32 K. In some Pb islands, this gap is also possible to be replaced by a peak feature (supplementary materials). These pseudogap and pseudopeak features have already been studied in Pb/Si system, explained as a modification to Fabry-Pérot mode due to the weaken of electron-phonon interaction below Debye energy [13]. Previous study on TiN has also found a pseudogap above $T_c$, which was explained as superconducting fluctuations [28]. According to our data, the pseudogap not only shows up above the $T_c$, but also coexists with the superconducting gap (see the dip features outside the superconducting gap in Fig. 4b). Therefore, the origin for pseudogap observed in our study is probably not the superconducting fluctuation.

The superconducting pairing strength is promoted by the electron-phonon coupling on Pb-STO interface. However, the development of Coulomb gap, originating from the enhanced electronic correlation in this size reduction system, simultaneously reduces the DOS near $E_F$. The competition between these two interactions might be the reason why $T_c$ is not enhanced. We demonstrate that the quantum size effect plays an important role to influence electronic correlation, which was usually ignored in two-dimensional superconducting systems in previous studies. Our finding sheds new light on understanding the interplay between superconductivity and complex electronic properties in low-dimensional quantum systems.

We thank T. Xiang, Y.Y. Wang, L. He, D. Liu, and W. Huang for helpful discussions. The experimental work was supported by the National Science Foundation (No. 11674191), Ministry of Science and Technology of China (No. 2016YFA0301002), and the Beijing Advanced Innovation Center for Future Chip (ICFC). W.L. was also supported by Beijing Young Talents Plan and the National Thousand-Young-Talents Program.

**Figures**

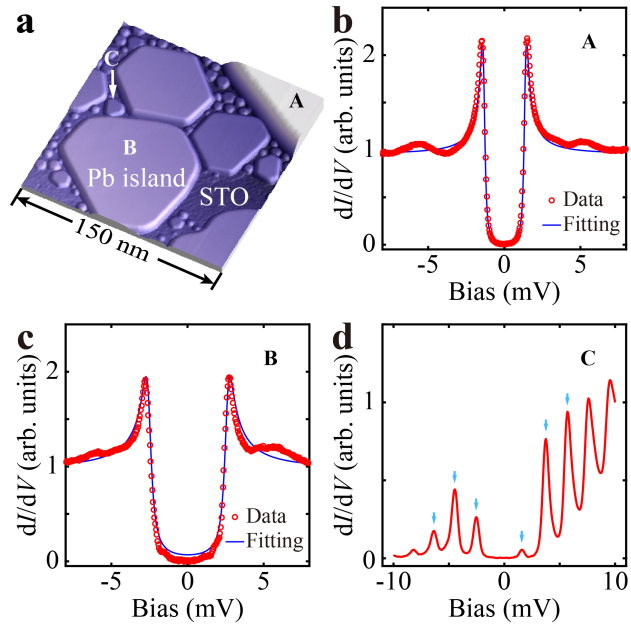

FIG. 1. (a) STM topographic image of Pb islands grown on STO(001) substrate (150 nm × 150 nm, $V_s$ = 2.0 V, $I_t$ = 20 pA). (b)(c) d$I$/d$V$ spectra taken on type A, B Pb islands grown on STO substrate, respectively (400 mK, $V_s$ = 10 mV, $I_t$ = 400 pA). The data are presented with red circles and the fitting results by Dynes model are in blue curves. Type A islands behave like bulk Pb and show a 1.36 meV superconducting gap. Type B islands show an enlarged 2.58 meV gap. (d) d$I$/d$V$ spectra taken on a type C Pb island (400 mK, $V_s$ = 10 mV, $I_t$ = 400 pA). The discrete peaks, denoted by cyan arrows, are signatures of Coulomb blockade.

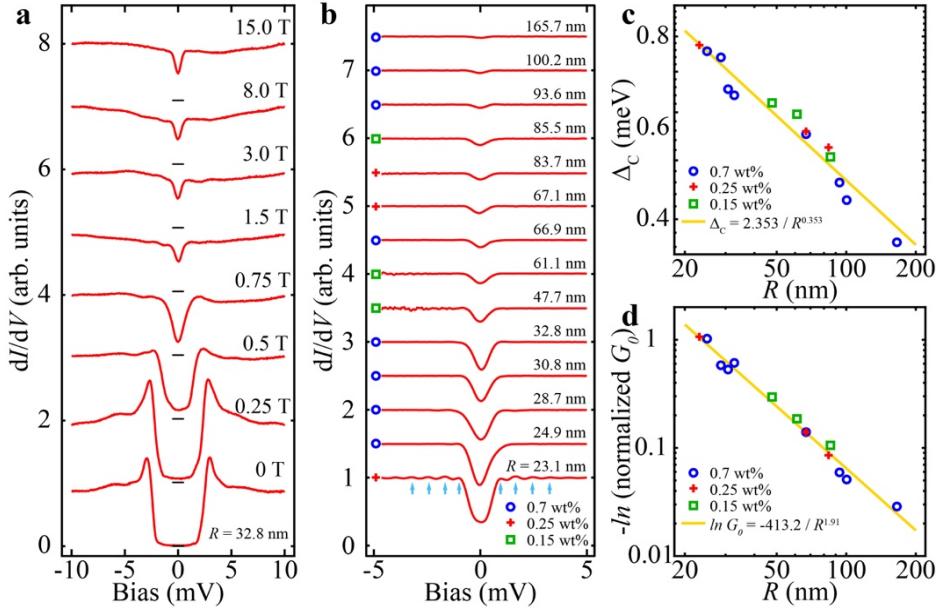

FIG. 2. (a) d$I$/d$V$ spectra taken on type B Pb islands ($R$ = 32.8 nm) under various magnetic fields. (400 mK, $V_s$ = 10 mV, $I_t$ = 300 pA). Upper critical field is enhanced, compared with that of bulk Pb. A residual gap still exists at high fields as superconductivity is totally suppressed. The magnetic insensitive residual gap is attributed to Coulomb gap. (b) Coulomb gaps of Pb islands with different areas (for details, see supplementary materials). The blue circles, red crosses and green squares denote 0.7 wt%, 0.25 wt% and 0.15 wt% Nb-doped substrates, respectively. The cyan arrows denote the Coulomb blockade features. (c) Double logarithmic scale plot of Coulomb gap size and lateral size of Pb island. The fitting result suggests $\Delta_C$ = 2.353/$R^{0.353}$. (d) Double logarithmic scale plot of negative logarithmic zero-energy density of states (-ln $G_0$) and lateral size of Pb island. $G_0$ exponentially decays with 1/Area, which obeys $G_0$ = $exp$ (-413.2/$R^{1.91}$).

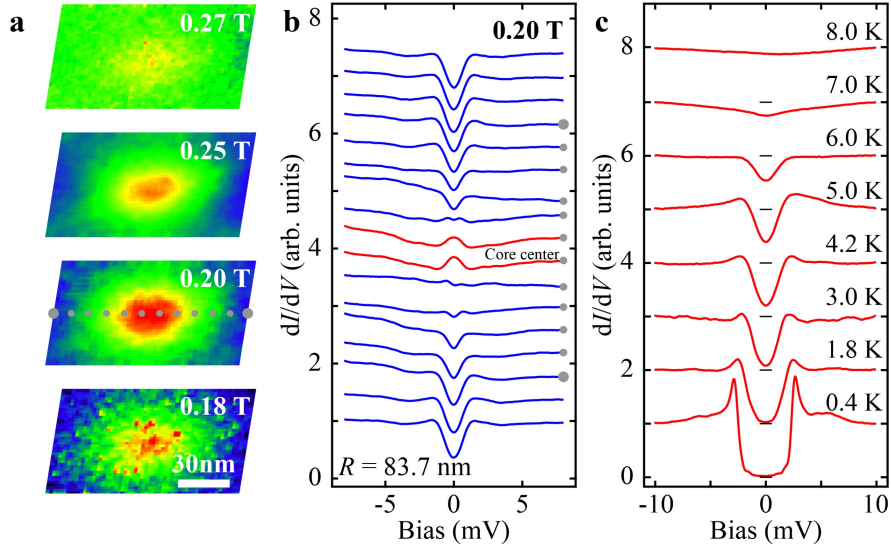

FIG. 3. (a) Zero-bias d$I$/d$V$ maps taken on a $R \sim 83.7$ nm island under different magnetic fields. The red color region, corresponding to higher conductance area, indicates the suppression of superconducting gap and emergence of a vortex. (b) A series of d$I$/d$V$ spectra taken across the vortex at 0.20 T (The corresponding positions are denoted by a grey dotted line in (a), $V_s = 10$ mV, $I_t = 200$ pA). A zero-bias peak is observed in the vortex core. (c) d$I$/d$V$ spectra taken on a $R = 30.8$ nm Pb island at various temperatures ($V_s = 10$ mV, $I_t = 200$ pA). Superconducting gap is almost suppressed at 7.0 K, which doesn't show significant enhancement compared with that of bulk Pb.

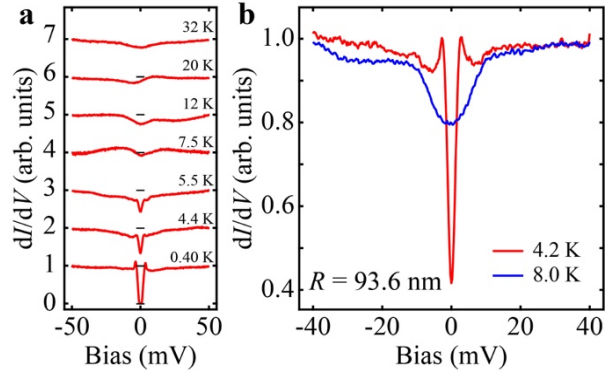

FIG. 4. (a) Larger-energy-range d$I$/d$V$ spectra taken on a $R \sim 100$ nm Pb island at various temperatures ($V_s = 50$ mV, $I_t = 200$ pA). When temperature is higher than $T_c$, a pseudogap persists near $E_F$. (b) Large range d$I$/d$V$ spectra taken on a $R = 93.6$ nm island ($V_s = 50$ mV, $I_t = 200$ pA). The red and blue curves show spectra taken at 4.2 K and 8.0 K, respectively. The pseudogap coexists with superconducting gap below $T_c$, as shown in the red curve.